\documentclass[aps,prl,preprint,superscriptaddress,showpacs]{revtex4-1}
\usepackage{epstopdf}
\usepackage{graphicx}  
\usepackage{dcolumn}   
\usepackage{bm}        
\usepackage{amssymb}
\usepackage[english]{babel}

\begin{document}


\title{A new Measurement of Thermal Conductivity of Iron at high pressures and temperatures}

\author{Pinku Saha}
\author{Goutam Dev Mukherjee}
\email [Corresponding Author:]{ goutamdev@iiserkol.ac.in}
\affiliation{Department of Physical Sciences, Indian Institute of Science Education and Research Kolkata, Mohanpur Campus, Mohanpur 741246, Nadia, West Bengal, India.}

\begin{abstract}
Thermal conductivity of the most abundant element in the planetary core, Iron (Fe) is measured up to  Earth's outer core pressure $\sim 120$  GPa. The measurements are carried out using the laser heated  diamond anvil cell facility, where the absorbed power by Fe metal foil is calculated using thermodynamical equation. The thermal conductivity of  $\gamma-Fe$ linearly increases up to a maximum experimental pressure 40 GPa. Thermal conductivity of $\epsilon-Fe$ measured by us shows a saturated  value $\sim$ 52 ($\pm$ 5) $Wm^{-1}K^{-1}$ in the pressure range 77 - 120 GPa. At different pressures temperature dependence of thermal conductivity show a sharp drop close to melting.  

\end{abstract}

\maketitle


\section{\label{sec:level1}Intruduction}

Convection in the liquid outer core of the Earth provides most of the energy to maintain self-sustained geodynamo, which results in a strong Earth’s magnetic field. Liquid outer core convection and and heat loss in the Earth surface strongly depend on the heat flux conducted at the outer most core of the planet. Knowledge of thermal conductivity of the core materials under extreme conditions of pressure and temperature is important to understand the evolution and dynamics of planets. Taking into account the effect of lattice thermal conductivity, theoretical calculations on Fe-Ni-Si alloy predicted a value 46 Wm$^{-1}$K$^{-1}$ at the core-mantle boundary [1] and a lower value (28-29 Wm$^{-1}$K$^{-1}$) was predicted by Stacey et al.[2]. First principles theoretical simulation studies of electrical resistivity of Iron demanded the thermal conductivity to be at the higher edge to 160-162 Wm$^{-1}$K$^{-1}$ at the inner core conditions [3]. First principles simulations predicted the thermal conductivity of Iron at 120 GPa and 2000 K to be 80 Wm$^{-1}$K$^{-1}$ [4]. Calculations using DFT predicted two times higher value than those by first principle simulations [5].


Four probe electrical resistivity measurements using multi-anvil and using Wiedemann-Franz-Law [6] estimated thermal conductivity of $\gamma-Fe$ in the range 40 - 100 Wm$^{-1}$K$^{-1}$ at the pressure and temperature range 5-7 GPa and 1000-1600 K, respectively. Electrical resistivity measurements by Gomi et al. [7] and Otha et al. [8] in DAC by four probe method also predicted the thermal conductivity value of Iron to a higher range, $>$ 90 Wm$^{-1}$K$^{-1}$ and 226 $^{+72}_{-31}$ Wm$^{-1}$K$^{-1}$, respectively at the Core-Mantle-Boundary (CMB). The electrical conductivity of iron measured by shock compression varied from 1.45$\times$10$^{4}$ $\Omega$$^{-1}$cm$^{-1}$ at 101 GPa and 2010 K, to 7.65$\times$10$^{3}$ $\Omega$$^{-1}$cm$^{-1}$  [9,10], which implied thermal conductivity values of 84 and 115 Wm$^{-1}$K$^{-1}$ at respective pressure and temperature points.  Theoretical measurements [3-5] and  indirect experimental measurements [6-10] show high thermal conductivity of Iron and its alloys at the outer core conditions, which implies a young inner core. Direct measurements of thermal conductivity of Iron  at the high pressures and temperatures are very rare. Thermal conductivity experiments using continuous wave (CW) IR laser, combined with finite-element numerical simulations by Konopkova et.al. estimated a value of 32 $\pm$ 7 Wm$^{-1}$K$^{-1}$  at 78 GPa and 2000K [11]. A very recent study [12] by the same group using a dynamically laser-heated diamond anvil cell showed a low value of thermal conductivity of Iron: (i) about 35 Wm$^{-1}$K$^{-1}$ at 48 GPa and 2000 K and (ii) in the range 18 - 44 Wm$^{-1}$K$^{-1}$ at about 130 GPa and 2000 K respectively. In both the experiments thermal conductivity was estimated from finite element simulation analysis and considering the reflectance of laser power from interfaces and thermal conductivity of pressure transmitting medium. All the above experiments and theoretical simulations showed a large variation in thermal conductivity of Fe at high pressures and temperatures. In fact Dobson raised the points that: (i) high thermal conductivity of Fe at CMB estimated by Otha et. al. may be an artifact due to the underestimate of heat loss through the electrical resistivity measurement electrodes and ; (ii) very low thermal conductivity estimated by Konopkova et. al. may due to the melting of Fe surfaces due to the very short laser pulses. These observations obviously opens up the controversies regarding the measurement of thermal conductivity values at the conditions of CMB. In the present work, we have carried out the measurement of the thermal conductivity of iron up to 120 GPa and 2000 K using single sided laser heated diamond anvil cell (LHDAC) under a steady state condition. We have calculated thermal conductivity from the direct temperature measurement of the sample surface and estimating the heat absorbed by the metal foil using thermodynamical equations. 

\section{\label{sec:level2}experiment and modeling}
Thermal conductivity measurements at high-pressures and high-temperatures were carried out using LHDAC facility using plate type diamond anvil cells (Almax-Boehler design). NaCl was used as pressure transmitting medium and as well as IR window. NaCl was kept in furnace at a temperature of 110$^{0}$ C for six hours to remove any trace of moisture. Steel gasket of thickness 225 $\mu$m was preindented to a thickness of 50 $\mu$m and a hole of diameter $\sim$ 110 $\mu$m was drilled at the center of diamond imprint of 300 $\mu$m culet. For 110 $\mu$m DAC culet the steel gasket preindented to a thickness of 40 $\mu$m and a hole of diameter $\sim$ 70 $\mu$m was drilled. In each run precompacted NaCl plates were used as the pressure transmitting medium. Thin plates of iron was made by compacting polycrystalline iron powder using a 300 ton hydraulic press. Thin pieces of Fe-plates of desired size (approximate diameter of about 50 - 80 $\mu$m) were cut to load them in the LHDAC. Pressure in the LHDAC before heating and after heating was determined using the ruby fluorescence method [14] by placing a few ruby chips (approximate sizes of about 3-4 $\mu$m) at the edge of the iron plate.

Heating was carried out using a diode-pumped Ytterbium ﬁber laser (YLR100-SM-AC-Y11) with central emission wavelength λ = 1.070 $\mu$m (maximum power 100 W). This wavelength is absorbed only by the Iron plate placed inside the sample chamber. The laser beam was focused  down to 16-18 $\mu$m in diameter to maintain localized heating. The temperature of the hot spot is controlled by monitoring the laser output power. For temperature measurements, the incandescent light from an area of 3 $\mu$m diameter from the metal surface was collected with SP150 Acton series spectrometer with back-illuminated PIXIS 100BR (pixel size:1340$\times$100) camera in the wavelength range 450-950 nm. The temperature measurement system was calibrated with the tungsten filament light source having known intensity vs wavelength distribution with temperature, bought from NPL, UK. The Plancks radiation function [15] was then fitted to the flatfield corrected spectrum:
 \begin{equation}
I(\lambda) = \frac{\varepsilon(P).C_1.\lambda^{-5}}{\exp\left\lbrace C_2/(\lambda.T)\right\rbrace-1},
\end{equation}
\noindent where, $I$ is the collected thermal radiation intensity, $\varepsilon(P)$ is the pressure dependent emissivity of the sample, $\lambda$ is the wavelength, $C_1 = 2hc^2$ and $C_2 = hc/k_B$ ($h$ is Planck's constant, $c$ is the speed of light and $k_B$ is Boltzman's constant). Fitting was done using $\varepsilon$ and $T$ as free fitting parameters, considering the grey body approximation.

Thermal conductivity of the iron plate was measured at steady state conditions assuming azimuthal and poloidal symmetry and following the equation:
\begin{equation}
k = \frac{Q.(r_1-r_2)}{4\pi.(T_1-T_2).r_1.r_2}
\end{equation}
where, $k$ is the thermal conductivity of the iron foil,$Q$ is the absorbed power in watt at the hotspot, $r_{1}$ is the radius of the hot spot, $r_{2}$ is the distance from the center of the hotspot where temperature is measured, $T_1$ is temperature at the hotspot and $T_{2}$ is the temperature at a distance $r_{2}$ from the center of hotspot. Temperature at different positions were estimated by translating the 50$\mu$m pin-hole attached to the spectrometer in a step size 8 $\mu$m. 




The absorbed power by the metal foil was calculated using thermodynamical equation:
\paragraph{}
\begin{equation}
Q=mc(T_1-T_0)\nu
\end{equation}
\noindent
where, $m$ is the mass of the hotspot = $\pi$r$^{2}$h$\rho$ (r is the radius of the hotspot, h is the thickness of the iron plate, $\rho$ is the density of the iron at respective pressure and is taken from Orson L. Anderson[16]), $c$ is the specific heat of iron at constant pressure [11,24], (T$_{1}$-T$_{0}$)= the temperature difference between hot spot and room temperature and $\nu$ is the modulated frequency of the 1.070 $\mu$m laser.

\section{\label{sec:level3}results and discussions}

Thermal conductivity of Fe is measured in fcc($\gamma$)-phase and hcp($\epsilon$)-phase by following the high pressure and the high temperature phase diagram of Fe [12]. We measured the temperature gradient on the sample surface by focusing the IR laser at center of the sample. A representative heating of the sample at 46 GPa is shown in Fig.1(a and b). The image of Fig.1(a) is taken under back illumination. Fig.1(b) shows enlarged image of the hot spot taken without back lighting and temperature is assigned at each step of the pinhole. The temperature gradient on the sample surface at four different pressure points (6, 31, 46 and 60 GPa) is shown in Fig.2. The pressure variation of emissivity of $\gamma$-Fe is shown in Fig.3.   Low value of emissivity at 10.3 GPa pressure and in temperature range 1820-1880K is attributed to close melting [18,19,20] of the iron. The error bar of emissivity in each pressure points is assigned by calculating the standard deviation. The sample pressure is estimated from the average value of the pressures measured before and after heating. Estimated emissivity ($\varepsilon$(P)) in $\gamma$-Fe is in  excellent agreement with data reported by Seagel et al.[17]. 


Temperature dependence thermal conductivity calculated from Equ. 2 estimating the absorbed power by Equ.(3) at lower pressures (6.4, 10.3 and 11.1 GPa) is shown in Fig.4. A sharp drop in the thermal conductivity is observed at temperatures $\sim$ 1760 K, 1940 K and 1960 K, respectively and is may be due to the hot spot temperature is close to the melting. The error-bar in temperature shown in Fig.4 are estimated from the uncertainty in temperature within hotspot in respective pressure points. The error-bar in the thermal conductivity of Fig.4 is calculated from temperature uncertainty within hotspot. Pressure dependence thermal conductivity of Fe in fcc($\gamma$)-phase , hcp($\epsilon$)-phase and their mixed phase are calculated by employing Eqn. 2 and 3 and are shown in Fig.5 along with other reported values. Filled symbols depict our data (blue filled triangle for $\gamma$-Fe and green filled triangle for $\epsilon$-Fe and orange filled triangle for mixed phase) and open symbols show the reported values in literature. Thermal conductivity of $\gamma$-Fe seems to increase linearly with pressure as shown by the dashed line. It shows a minimum value 30 Wm$^{-1}$K$^{-1}$ at pressure 6.4 GPa around 1600 K to a maximum value 50 Wm$^{-1}$K$^{-1}$ at pressure 39 GPa around 1900 K. Thermal conductivity of the mixed phase (hot spot at $\gamma$ phase and edge at $\epsilon$ phase as per the phase diagram [12]) has a value in the range 19-30 Wm$^{-1}$K$^{-1}$ in the pressure and temperature range of 21.6 to 39 GPa and 1400 to 1700 K, respectively. The thermal conductivity measured above 46.2 GPa is from single phase of  $\epsilon$-Fe and has value in the range 31 to 53 Wm$^{-1}$K$^{-1}$. Thermal conductivity above 77 GPa shows saturation at a value about 52 Wm$^{-1}$K$^{-1}$.

Our low pressure $k$ value for $\gamma$-Fe is found to be close to the ambient pressure value reported by  Ho et al. [23] indicating accurate measurements of thermal conductivity using our new method. In $\gamma$-phase $k(P)$ seems to be linear with respect to pressure and are close to the values reported by Konopkova et al [12]. In $\epsilon$-phase $k(P)$ seems to follow linear relation with pressure below about 70 GPa. Above it seems to saturate. In this phase our measured values are slightly more than the reported data of Konopkova et al [12]., however well within their error bar.

\section{\label{sec:level4}conclusions}
To summarize, a new technique has been employed to calculate the thermal conductivity in laser heated DAC. In this technique thermodynamical equation has been used to calculate absorbed power by the metal plate.The thermal conductivity of iron is measured up to Earth's outer core pressure $\sim$ 120 GPa. Value of thermal conductivity of $\gamma$ iron is higher than that of $\epsilon$ iron at the equivalent pressure point. A strong dependence of $k$ values of $\gamma$ Fe on pressure is observed and a saturated values of $k$ is observed in $\epsilon$ iron in the pressure range 77- 120 GPa.

\begin{figure}
\includegraphics[width=13cm,height=7cm]{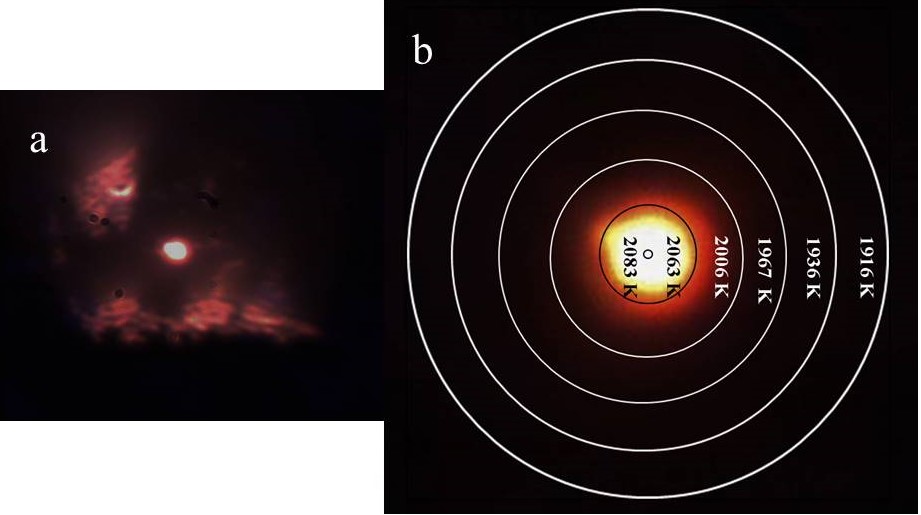}

\caption{\label{fig1} (a) The image of hotspot about  diameter   16 – 18 $\mu$m on Iron at 46 GPa under transmitting light. (b) The magnified image of  hotspot  on Iron at 46 GPa with radial  temperature distribution. }
\end{figure}
\begin{figure}
\includegraphics[width=12cm,height=15cm]{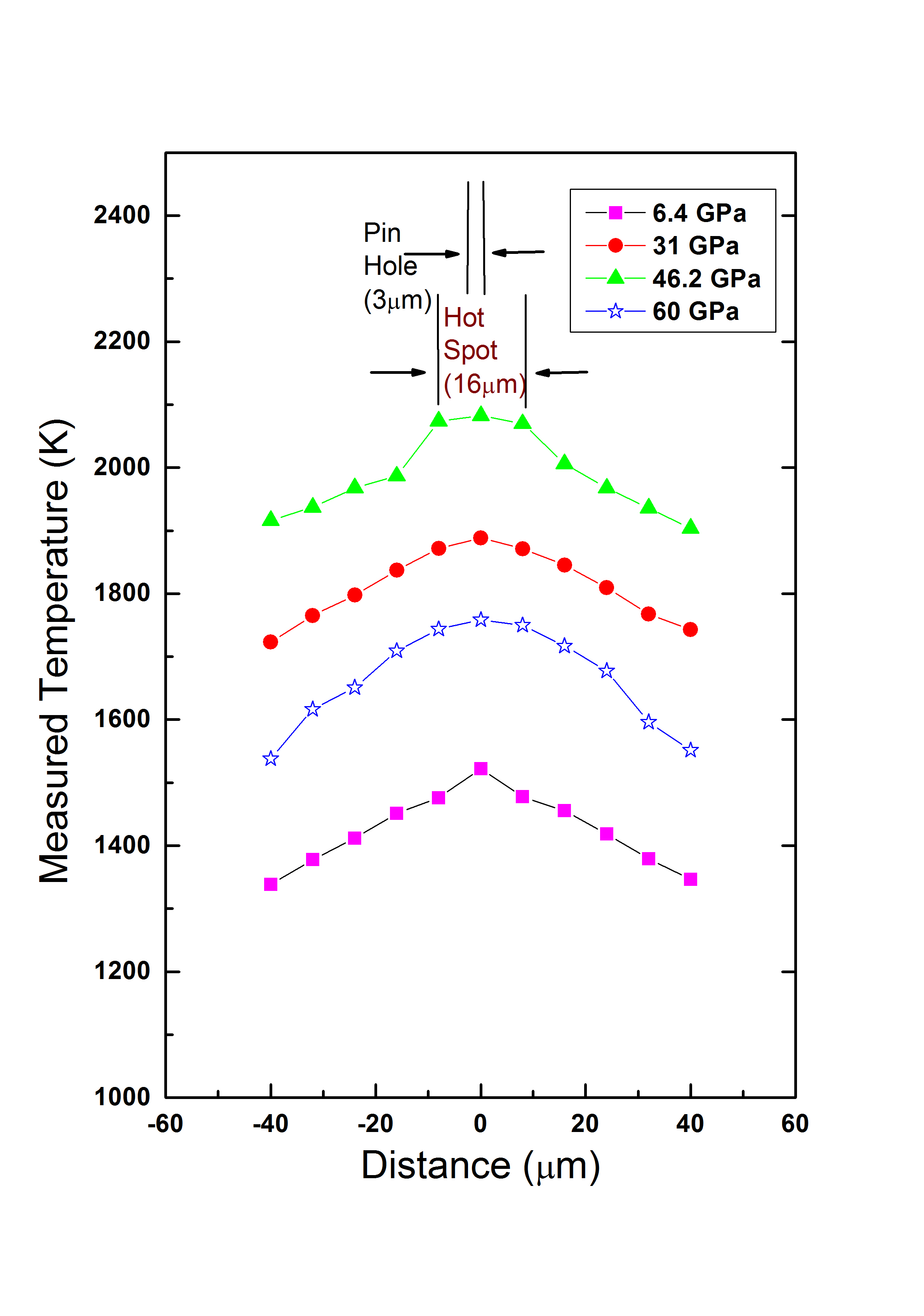}

\caption{\label{fig2} Temperature gradient on iron foil heated at different pressures. Temperature were measured from areas with 3$\mu$m diameter.}
\end{figure}
\begin{figure}
\includegraphics[width=12cm,height=15cm]{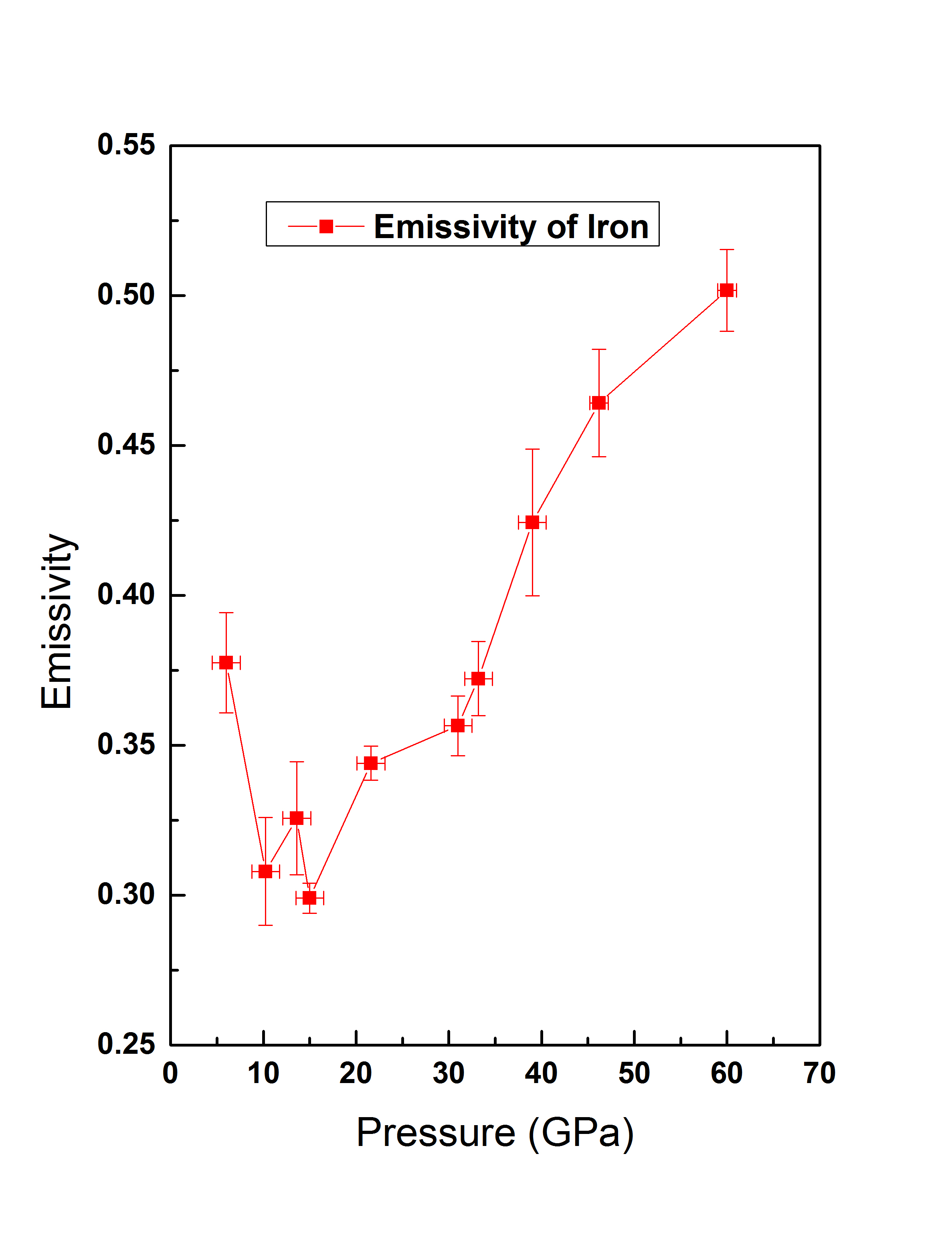}
\caption{\label{fig3} Pressure dependence emissivity of iron estimated from the two parameter fitting of Planck's radiation function.}
\end{figure}
\begin{figure}
\includegraphics[width=12cm,height=15cm]{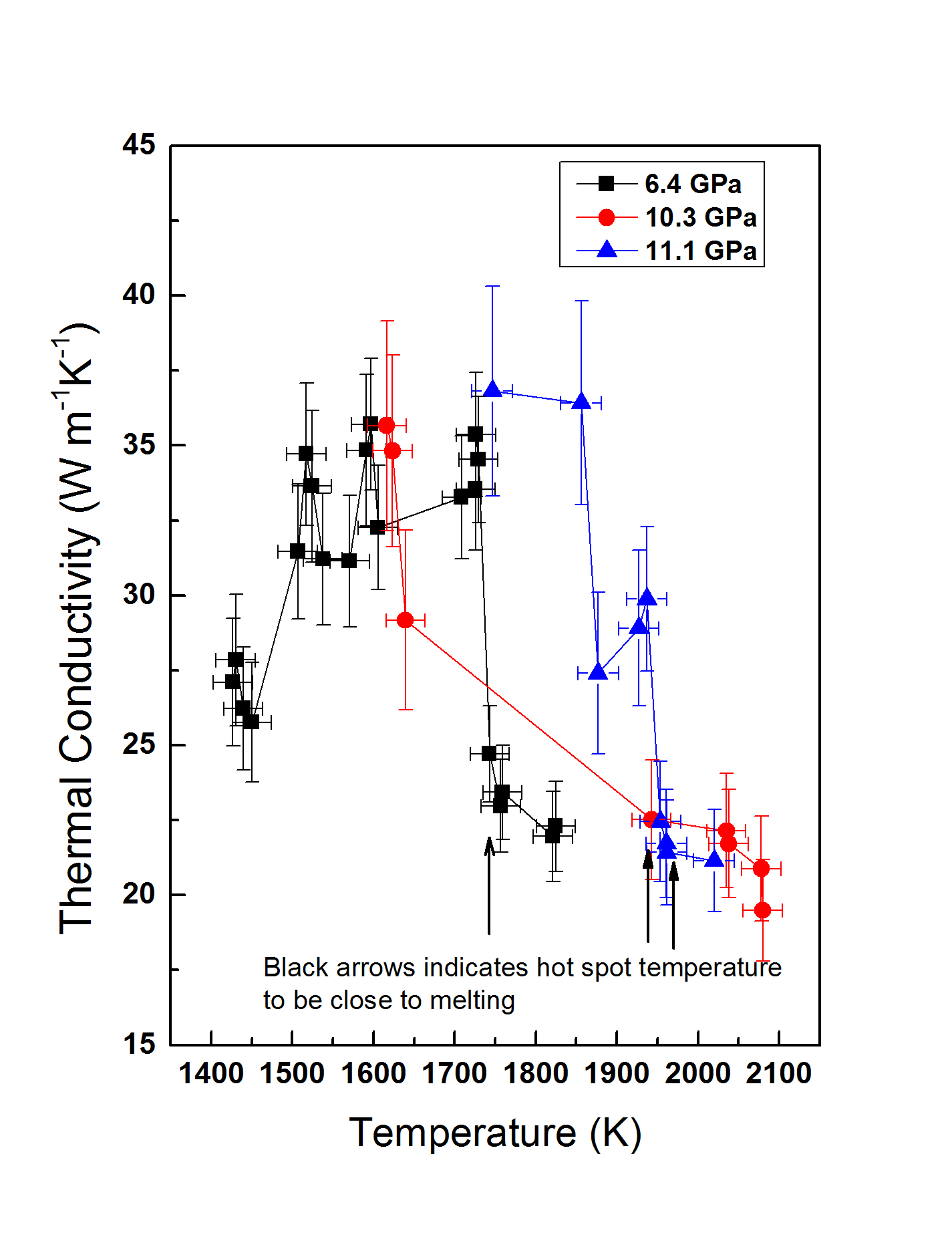}

\caption{\label{fig4} Temperature dependent thermal conductivity of iron at pressures 6.4, 10.3 and 11.1 GPa. The sharp drop in the thermal conductivity values indicated by black arrow may be due to hot spot temperature is close to melting of iron.}
\end{figure}
\begin{figure}
\includegraphics[width=12cm,height=15cm]{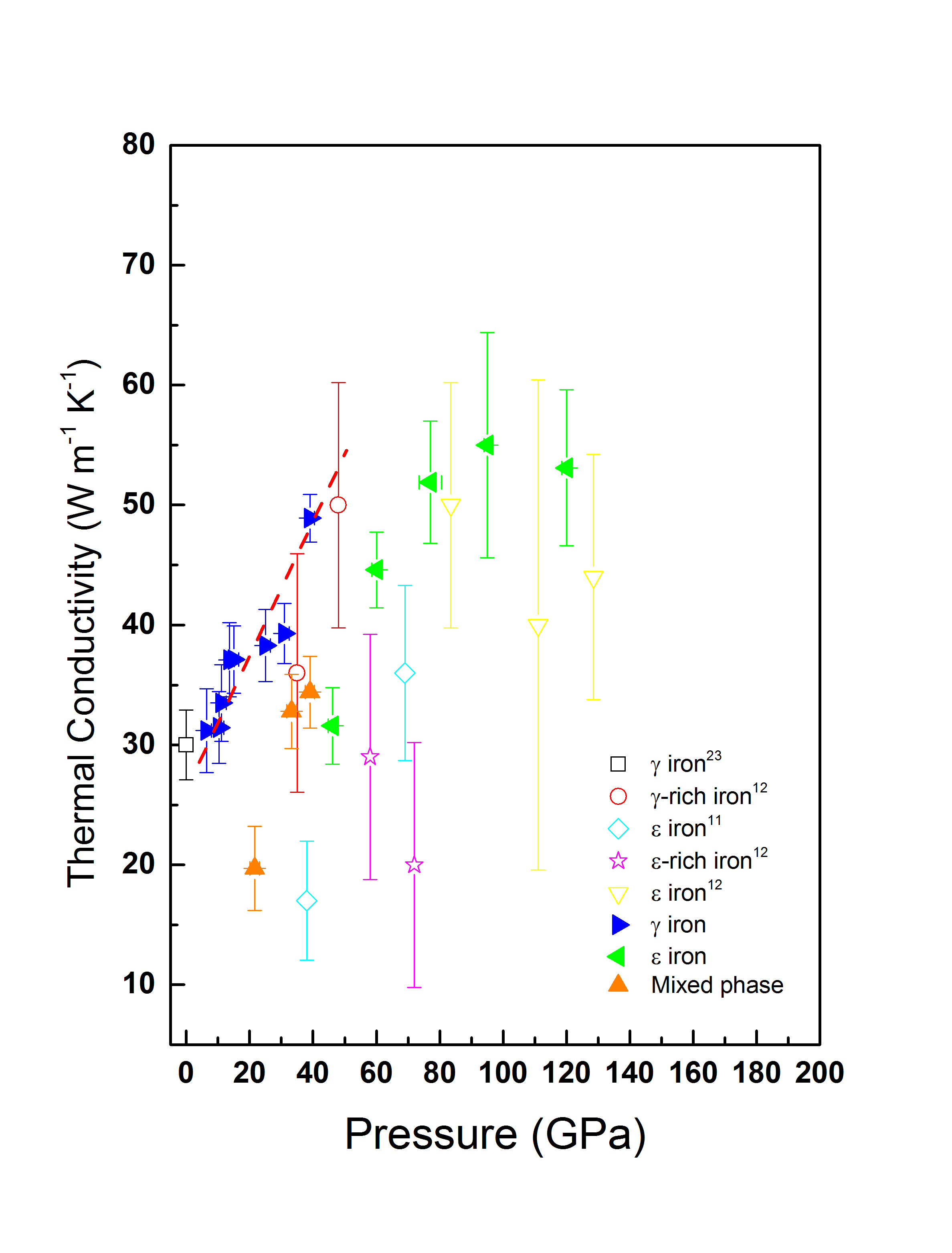}

\caption{\label{fig5} Pressure dependent thermal conductivity of iron. Filled triangle are the results of our study. Thermal conductivity value of $\gamma$ iron is shown as blue triangle, $\epsilon$ iron by green triangle and mixed phase of $\gamma$ and $\epsilon$ iron by orange triangle.}
\end{figure}
\end{document}